\documentclass[aps,prl,twocolumn,showpacs,groupedaddress]{revtex4}
\usepackage{amsmath}
\usepackage{amssymb}
\usepackage{epsfig}
\usepackage{graphicx,amsmath}
\usepackage{calc}
\usepackage{bm}
\begin{document}
\title{Ultra-high mobility two-dimensional electron gas in a SiGe/Si/SiGe quantum well}
\author{M.~Yu. Melnikov\footnote{e-mail: melnikov@issp.ac.ru}, A.~A. Shashkin, and V.~T. Dolgopolov}
\affiliation{Institute of Solid State Physics, Chernogolovka, Moscow District 142432, Russia}
\author{S.-H. Huang and C.~W. Liu\footnote{e-mail: chee@cc.ee.ntu.edu.tw}}
\affiliation{Department of Electrical Engineering and Graduate Institute of Electronics Engineering, National Taiwan University, Taipei 106, Taiwan, and\\ National Nano Device Laboratories, Hsinchu 300, Taiwan}
\author{S.~V. Kravchenko}
\affiliation{Physics Department, Northeastern University, Boston, Massachusetts 02115, USA}
\date{\today}

\begin{abstract}
We report the observation of an electron gas in a SiGe/Si/SiGe quantum well with maximum mobility up to
240 m$^2$/Vs, which is noticeably higher than previously reported results in silicon-based structures. Using SiO, rather than Al$_2$O$_3$, as an insulator, we obtain strongly reduced threshold voltages close to zero. In addition to the predominantly small-angle scattering well known in the high-mobility heterostructures, the observed linear temperature dependence of the conductivity reveals the presence of a short-range random potential.
\end{abstract}

\maketitle

During last two decades, a big progress has been made in studying the behavior of strongly correlated electrons confined in two dimensional (2D) quantum wells. The metal-insulator transition, discovered initially in low-disorder silicon metal-oxide-semiconductor field-effect transistors (MOSFETs) and subsequently observed in many other strongly correlated two-dimensional systems \cite{abrahams2001}, has attracted a great deal of attention and is still not completely understood \cite{spivak2010}. It is generally agreed that for the zero-magnetic-field metallic state to exist in 2D, strong interactions between carriers are needed. Other effects of strong correlations include greatly enhanced spin susceptibility \cite{shashkin2001} and the effective mass \cite{shashkin2002}. The strength of interactions is usually characterized by the ratio of Coulomb ($E_c$) and Fermi ($E_F$) energies that is related to the dimensionless Wigner-Seitz radius $r_s$:
\begin{equation}
\frac{E_C}{E_F}=\frac{e^2n_vm}{\epsilon \hbar^2(\pi n_s)^{1/2}}=n_vr_s,
\end{equation}
where $n_v$ is the valley degeneracy, $m$ is the effective mass, $\epsilon$ is the dielectric constant, and $n_s$ is the electron density. The 2D electron system in silicon turned out to be a very convenient object for studies of the strongly correlated regime due to relatively high effective band mass ($0.19\, m_e$ compared to $0.067\, m_e$ in n-type GaAs/AlGaAs heterostructures) and the existence of two almost degenerate valleys in the spectrum ($n_v=2$), which further enhances the correlation effects \cite{punnoose2005}. However, even in the highest quality Si MOSFETs, the maximum electron mobilities did not exceed 3 m$^2$/Vs, and the effects of the residual disorder mask the effects of interactions on the insulating side of the metal-insulator transition, where the formation of the Wigner crystal is expected. This calls for a new high-mobility silicon-based 2D system to be developed.

In this Letter, we report the observation of the 2D electron gas in extremely low-disorder SiGe/Si/SiGe quantum wells with mobilities up to 240 m$^2$/Vs, {\it i.e.}, almost two orders of magnitude higher than those in the best silicon MOSFETs and noticeably higher than in previously fabricated SiGe heterostructures \cite{schaffler1997,lai2004,dolgopolov2003}. The main improvements to the mobility and threshold voltage, compared to previously fabricated SiGe/Si/SiGe quantum wells, come from using an ultrahigh-vacuum chemical-vapor-deposition (UHVCVD) instead of MBE and employing SiO, rather than Al$_2$O$_3$ \cite{lu2009,erratum,huang2012}, as an insulator. Our samples are also characterized by an extremely high homogeneity of the electron density which is crucial for measurements in the low-density limit where correlations become important. An-order-of-magnitude difference between transport and quantum relaxation times points to a predominantly small-angle scattering well known in the high-mobility heterostructures, although the observed linear temperature dependence of the conductivity indicates also the presence of a short-range random potential.

\begin{figure}
\includegraphics[width=8cm]{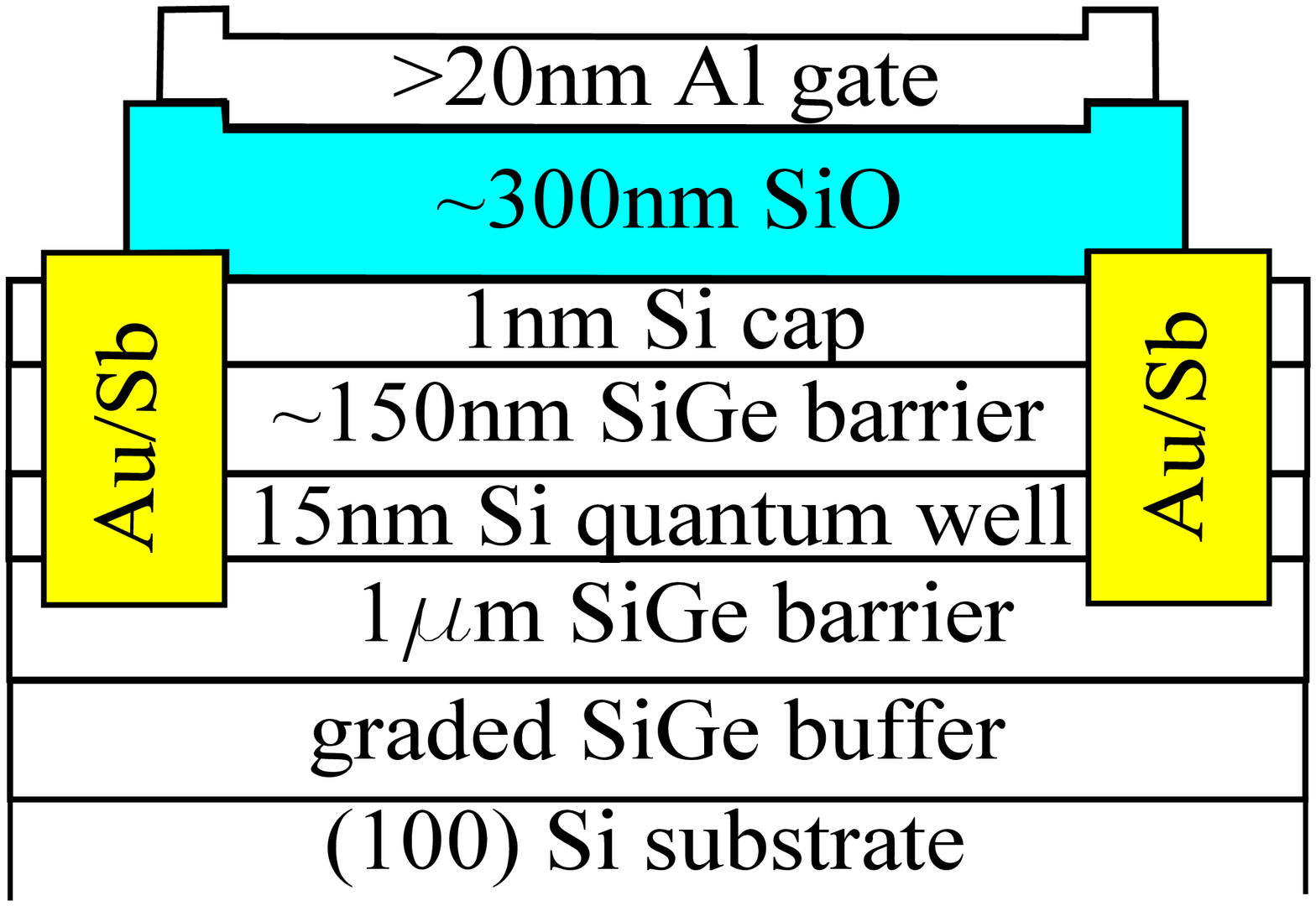}
\caption{\label{fig:scheme} (color online). Band diagram of the sample.}
\end{figure}

To make a sample, we used a SiGe/Si/SiGe quantum well grown in UHVCVD (for details, see Refs.~\cite{lu2009,erratum,huang2012}). Approximately 15 nm wide silicon quantum well is sandwiched between SiGe potential barriers (Fig.~\ref{fig:scheme}). The Ge content of the SiGe relaxed buffer layer is $\sim18$\% and the silicon quantum well layer is fully strained. The samples were patterned in Hall-bar shapes using standard photo-lithography (for details, see Ref.~\cite{melnikov2014}) on two different pieces, SiGe1 and SiGe2, of the same wafer. As the first step, electric contacts to the 2D layer were made. They consisted of AuSb alloy, deposited in a thermal evaporator in vacuum and annealed. Then, approximately 300 nm thick layer of SiO was deposited in a thermal evaporator \cite{blevis1963} and a $>20$ nm thick Al gate was deposited on top of SiO. No mesa etching was used, and the 2D electron gas was created in a way similar to silicon MOSFETs. The fabrication procedures of SiGe1 and SiGe2 differed significantly only in the way of how the electric contacts to the 2D layer were made. In case of SiGe2, after depositing approximately 350 nm Au$_{0.99}$Sb$_{0.01}$, the contacts were annealed for 5 minutes in N$_2$ atmosphere at 440$^\circ$C (the procedure used in Refs.~\cite{lu2009,erratum,huang2012}). In case of SiGe1, first about 30 nm of Sb and then 230 nm of Au were deposited, and then annealing was made by an electric spark which was produced by touching the contact with a metallic needle while the second metallic needle, connected with the first one by a charged capacitor, was pressed to the Au/Sb surface. Several of such annealing were done for each of the contact pads along the edge, where the Al gate was subsequently deposited.

When a positive voltage $V_g>V_{th}\approx0$ is applied to the gate, an $\approx15$ nm wide electron system is formed in a Si (100) quantum well approximately 150 nm below the SiO layer. It is expected that the properties of such a system (the band electron mass, $g$-factor, two-valley spectrum) are identical with those of the 2D system in Si MOSFETs, with the exception of the characteristic energy of the electron-electron interactions. The latter is expected to be somewhat weaker than that in Si MOSFETs because of a greater average dielectric constant in SiGe/Si/SiGe ($\sim12$ compared to 7.7 in Si MOSFETs).

\begin{figure}
\includegraphics[width=8cm]{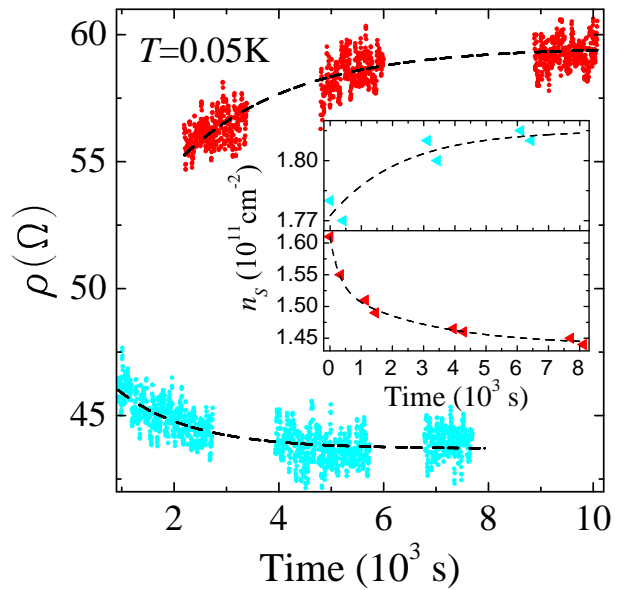}
\caption{\label{fig:R(t)} (color online). Resistivity of SiGe1-II as a function of time after decreasing $V_g$ to 3.2 V (red) and after increasing it to 4.2 V (blue). The inset shows corresponding density dependence on time. The dashed lines are guides to the eye.}
\end{figure}

We have studied four samples of the same geometry, two of SiGe1 type and two of SiGe2 type, in an Oxford TLM-400 dilution refrigerator in a temperature range 0.05 -- 1.2 K. To measure the resistance, a standard four-terminal lock-in technique was used in a frequency range 1 -- 11 Hz; the applied currents varied in the range 0.5 -- 4 nA. The electron density was determined from the low-field Shubnikov-de~Haas oscillations (see below); electron densities determined from the Hall effect were found to be the same within 10\%. The ohmic contacts to the 2D layer became highly resistive already at relatively high electron densities $n_s\lesssim1.2$ -- $1.6\times10^{11}$ cm$^{-2}$ (depending on the sample) and disappeared altogether at yet lower densities. We have found out that after an illumination of the sample by an infra-red LED, the contacts greatly improved and the electron densities down to $0.2\times10^{11}$ cm$^{-2}$ could be reached. Most of the results presented in this Letter have been obtained in the dark regime; however, some measurements on two highest-mobility samples (SiGe2-I and SiGe2-II) have been performed after the infra-red illumination. Since the contact resistance was often in the range between 0.5 and 100 k$\Omega$, a preamplifier with the input resistance of 100 M$\Omega$ was used to minimize the effect of the contact resistance. The second problem was that it took rather long time for the electron density to stabilize after the gate voltage was changed: about two hours for SiGe1 type of samples and less than half an hour for SiGe2 type (see Fig.~\ref{fig:R(t)}). The electron density was carefully monitored during the measurements. In both types of samples, if a high density above $2.4\times10^{11}$~cm$^{-2}$ was initially set by the gate voltage, after a few hours it would ultimately reduce to $n_s\approx2.4\times10^{11}$ cm$^{-2}$. The same effect was reported in Ref.~\cite{lu2011} where a similar maximum electron density of $2.7\times10^{11}$ cm$^{-2}$ was reported. The saturation of the electron density was explained by a tunneling of the electrons through the SiGe barrier at high gate voltages.

\begin{figure}
\includegraphics[width=8cm]{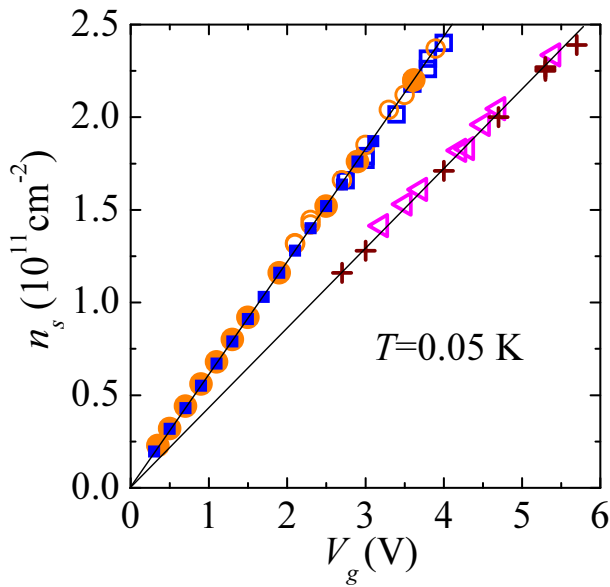}
\caption{\label{fig:n(V)} (color online). Electron density as a function of the gate voltage for four samples.  For each of the samples, the data were obtained in several cool-downs from room to helium temperature. The solid lines are linear fits to the data. Empty squares: SiGe2-I in the dark regime; solid squares: SiGe2-I after the illumination; empty circles: SiGe2-II in the dark regime; solid circles: SiGe2-II after the illumination; crosses and triangles: SiGe1-I and SiGe1-II, correspondingly, in the dark regime.}
\end{figure}

The threshold voltage $V_{th}\approx 0$ was determined by extrapolating the linear dependence of the stabilized electron density on the gate voltage, as shown in Fig.~\ref{fig:n(V)}. (The infra-red illumination of the sample did not affect the threshold within the experimental accuracy.) In contrast, a rather high threshold voltage $V_g=5.25$ V was reported in Refs.~\cite{lu2009,erratum,huang2012} where Al$_2$O$_3$ was used as a dielectric between the structure and the gate. According to the authors, this might be due to the influence of the interface between the Al$_2$O$_3$ layer and the heterostructure. It is worth noting that zero threshold voltage in our samples indicates much higher quality of the interface.

\begin{figure}
\includegraphics[width=8.5cm]{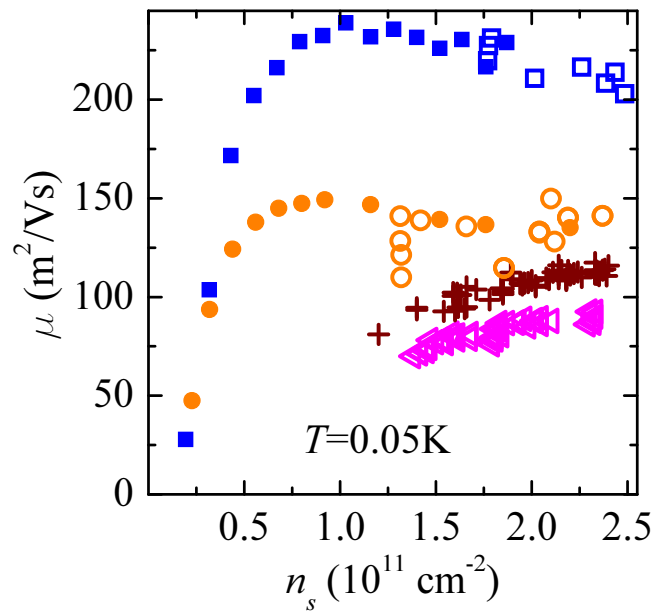}
\caption{\label{fig:mobility(n_s)} (color online). Mobility as a function of $n_s$ for four samples at $T=50$~mK. The symbols are identical to those in Fig.~\ref{fig:n(V)}.}
\end{figure}

The electron mobility $\mu(n_s)$ in zero magnetic field $B=0$ for four samples is shown in Fig.~\ref{fig:mobility(n_s)}. For each of the samples, the data were obtained in several cool-downs from room to helium temperature, which resulted in appreciable scattering of the data obtained without illumination. At $T=0.05$~K and $n_s=1.0$ -- $2.4\times10^{11}$ cm$^{-2}$, the maximum electron mobility varied between 90 and 240 m$^2$/Vs for the four samples studied. The mobility variation in the samples is determined by the Ge content, quantum well width, and defect distributions in the wafer.

\begin{figure}
\includegraphics[width=8cm]{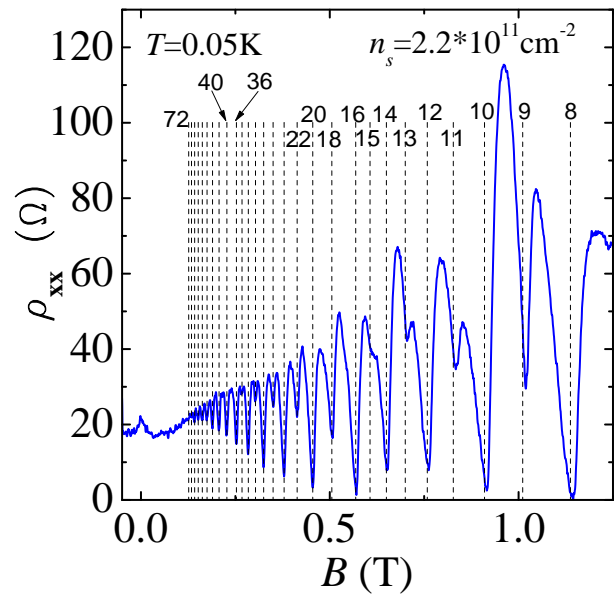}
\caption{\label{fig:R(B)} (color online). The longitudinal resistivity as a function of the perpendicular magnetic field for SiGe2-II. The filling factors $\nu=n_s hc/eB$, corresponding to the minima of the Shubnikov-de~Haas oscillations, are marked by vertical dashed lines. The data were obtained after the illumination.}
\end{figure}

The longitudinal resistivity $\rho_{xx}$ as a function of the perpendicular magnetic field $B$ is shown in Fig.~\ref{fig:R(B)}. At low magnetic fields, the filling factors $\nu=n_s hc/eB$, corresponding to the minima of the Shubnikov-de~Haas oscillations, are factors of 4 in agreement with the existence of the two-fold spin and valley degeneracies of the Landau levels; at magnetic fields above approximately 0.25~T, the spin splittings become resolved (filling factors $4n+2$, where $n$ is an integer), and at $B\gtrsim0.6$~T, resistance minima at odd filling factors, corresponding to the valley splitting, are visible. The quantum oscillations start at a magnetic field of about 0.1~T. This allows one to estimate the ``quantum'' mobility to be of order 10 m$^2$/Vs. However, the value of the mobility, calculated from the conductivity data, is an order of magnitude higher. It means that the transport relaxation time in our samples is an order of magnitude longer than the quantum time $\tau_q$ responsible for the width of the Landau levels. Similar an-order-of-magnitude difference between transport and quantum relaxation times has been observed on all four samples at all densities. This points to the predominantly small-angle scattering well known in the high-mobility heterostructures.

\begin{figure}
\includegraphics[width=8.5cm]{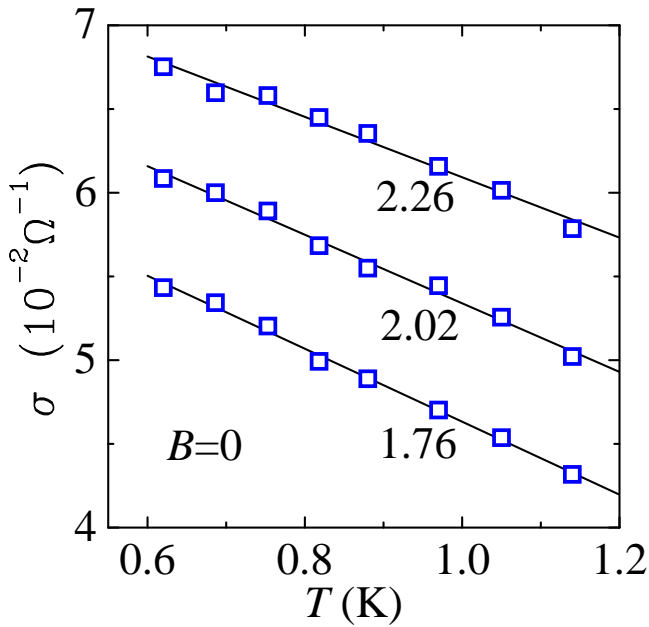}
\caption{\label{fig:sigma(T)} (color online). The temperature dependence of the conductivity at different electron densities indicated in units of $10^{11}$~cm$^{-2}$ for SiGe2-I. The solid lines are linear fits. The data were obtained in the dark regime.}
\end{figure}

The behavior of the conductivity $\sigma$ with temperature in the temperature range where the $\sigma(T)$ dependence is linear is displayed in Fig.~\ref{fig:sigma(T)}. The metallic temperature dependence of $\sigma$ ($d\sigma/dT<0$) is typical of 2D systems with low disorder \cite{shashkin2005}. The observed linear-in-$T$ correction to conductivity indicates also the presence of backscattering in the 2D electron system, i.e., the presence of a short-range random potential \cite{gold1986}. With the correction in the form $-\alpha\sigma_0k_BT/E_F$ (where $\alpha$ is a coefficient and $\sigma_0$ is a linear extrapolation of the conductivity to $T=0$) \cite{gold1986,zala2001} and the renormalized mass taken from Ref.~\cite{melnikov2014}, the value of $\alpha\approx2.7$ is determined to be about the same as the one obtained in silicon MOSFETs \cite{shashkin2004}.

Finally, we mention that the electron density in our samples is extremely homogeneous as inferred from the Shubnikov-de~Haas oscillations being resolved up to filling factor of 72 (down to $\approx0.1$~T) and the fact that the measured capacitance of the samples (e.g., 15.6 pF for SiGe2-I) practically coincides with the one calculated (15.5 pF) from the data shown in Fig.~\ref{fig:n(V)} and the known area of the samples.

In conclusion, we have manufactured extremely uniform SiGe/Si/SiGe quantum wells with maximum mobilities up to 240 m$^2$/Vs. This is noticeably higher than the previously reported mobilities reached in SiGe heterostructures and almost two orders of magnitude higher than the mobilities reached in the best Si MOSFETs.

We gratefully acknowledge discussions with G. M. Minkov and V. N. Zverev. This work was supported by RFBR 15-02-03537 and 13-02-00095, RAS, and the Russian Ministry of Sciences. NTU group was supported by Ministry of Science and Technology (103-2622-E-002 -031). SVK was supported by NSF Grant 1309337 and BSF Grant 2012210.

\end{document}